\documentclass[
reprint,
amsmath,
amssymb,
pra,
aps]{revtex4-2}

\usepackage{tensor}
\usepackage{xcolor}
\usepackage{physics}
\usepackage{graphicx}
\usepackage{todonotes}
\usepackage[colorlinks=true,linkcolor=blue,citecolor=blue,urlcolor=blue]{hyperref}

\newcommand{\Z}{\mathbb{Z}}

\begin{document}

\title{Non-renormalisation of coupling constants \texorpdfstring{\\}{} from categorical symmetries in two dimensions}

\author{Guillermo Arias-Tamargo}
\author{Chris Hull}
\author{Maxwell L.~Vel\'{a}squez Cotini Hutt}
\affiliation{Abdus Salam Centre for Theoretical Physics, The Blackett Laboratory, Imperial College London, Prince Consort Road, London, SW7 2AZ, UK}

\begin{abstract}
We study the role of categorical symmetries in constraining the renormalisation of couplings in two-dimensional non-linear sigma models with Wess-Zumino term. A large class of these theories admit self-duality symmetries associated with discrete gauging and T-duality. They are generically non-conformal, but we argue that a particular coupling is protected from quantum corrections by the categorical symmetry. We give strong evidence for this claim by showing that the $\beta$-function for this coupling vanishes to 2-loop order if and only if this symmetry is present. Furthermore, in cases where the target space is a group manifold, the non-renormalisation result can be proven to hold non-perturbatively. 
\begin{description}
\item[E-mail] g.arias-tamargo@imperial.ac.uk, c.hull@imperial.ac.uk, m.hutt22@imperial.ac.uk
\end{description}
\end{abstract}

\maketitle

\section{Introduction}\label{sec:intro}

An important role of symmetries is to  constrain the behaviour of complicated physical systems.
In the context of Quantum Field Theory (QFT), the aim is typically to understand the dynamics of the fields in the strong coupling regime. Global symmetries are one tool which can sometimes be exploited to obtain quantitative predictions for such systems. A prominent example is supersymmetry, which facilitates the exact computation of the low energy effective field theory of asymptotically free supersymmetric gauge theories. In other cases, when such powerful results are generically out of reach, one can often still extract conclusions about the qualitative behaviour of the system, for example through 't Hooft anomaly matching. 

The modern paradigm of generalised global symmetries \cite{Gaiotto:2014kfa} has shown that there are many more such constraints than had been previously realised. We will focus here on so-called \emph{categorical} or \emph{non-invertible} symmetries \footnote{See, e.g.~\cite{Bhardwaj:2023kri, Schafer-Nameki:2023jdn, McGreevy:2022oyu, Brennan:2023mmt, Costa:2024wks, Shao:2023gho} for reviews.}. Examples of their applications to QFTs in two dimensions from various approaches include \cite{Chang:2018iay,Thorngren:2021yso,Komargodski:2020mxz,Bhardwaj:2023idu,Bhardwaj:2023fca,Damia:2024kyt,Copetti:2024rqj,Copetti:2024dcz,Pace:2024acq,Cordova:2024nux,Cordova:2024goh}. 
A class of categorical symmetries which give rise to particularly strong constraints are known as \emph{self-duality symmetries} \cite{ChoiCordovaHsinLam2021}, which were originally studied in two dimensional conformal field theories \cite{Chang:2018iay,Thorngren:2021yso,Oshikawa:1996ww,Oshikawa:1996dj,Petkova:2000ip,Fuchs:2002cm,Fuchs:2003id,Fuchs:2004dz,Fuchs:2004xi,Frohlich:2004ef,Frohlich:2006ch,Fuchs:2007tx,Bachas:2012bj,Bachas:2013ora,Niro:2022ctq,Bharadwaj:2024gpj,Damia2024,Argurio:2024ewp}. The goal of the present work is to investigate the consequences of these symmetries in non-conformal Non-Linear Sigma Models (NLSMs) in two dimensions. 

A Non-Linear Sigma Model with Wess-Zumino (WZ) term is a theory of maps $\Phi:W\to M$  with action
\begin{align}
\begin{split}\label{eq:action}
    S &= \frac{1}{2\pi\alpha'} \bigg( \frac{1}{2}\int_W g_{ij}\, dX^i\wedge\star dX^j \\
    &\qquad\qquad+ \frac{1}{3!}\int_V H_{ijk}\, dX^i\wedge dX^j\wedge dX^k \bigg)\,,
\end{split}
\end{align}
where $g$ and $H$ are a metric and a closed 3-form on the target space $M$ respectively, the fields $X^i$ are coordinates on $M$ ($i=1,\dots,N=\dim M$), and $W$ is a closed two-dimensional worldsheet which is  the boundary of a 3-manifold $V$. This is an interacting QFT, with various non-linear couplings appearing through $g_{ij}(X)$ and $H_{ijk}(X)$. As a result, the geometry of the target space itself is subject to Renormalisation Group (RG) flow. In some cases, the NLSM can have a coupling with respect to which the theory is asymptotically free and strongly coupled in the IR.

It has recently been observed that a wide class of these models have a categorical symmetry associated to self-dualities under discrete gauging \cite{Arias-Tamargo:2025xdd,Arias-Tamargo:2025fhv}. This symmetry is present when the target space $M$ has a compact isometry generated by a Killing vector $k$ such that
\begin{equation}\label{eq:Lie_derivatives}
    \mathcal{L}_k g = 0\,,\quad \mathcal{L}_k H=0\,,
\end{equation}
and its norm squared $G=g_{ij}k^i k^j$ satisfies
\begin{align}\label{eq:self_duality_radius}
    G=\frac{p}{2\pi q} 
\end{align}
with $p,q$ coprime integers. (There are further conditions on the NLSM to admit the self-duality defect which we discuss later).

Consider a NLSM whose couplings are tuned in the UV so that they satisfy \eqref{eq:self_duality_radius}, so the theory has the categorical symmetry. Group-like global symmetries are not broken by RG flow unless an explicit symmetry breaking deformation is introduced, and the expectation is that this should apply equally to categorical symmetries \cite{Chang:2018iay}. In the present context, this implies that the relation \eqref{eq:self_duality_radius} should be preserved under RG flow, or else the categorical symmetry would be broken. In what follows, we will check that this is indeed the case to lowest orders in perturbation theory, and give examples where it holds at the non-perturbative level. In other words, we will argue that the coupling $G$, at the specific values $\eqref{eq:self_duality_radius}$, is protected from all quantum corrections by the categorical symmetry. Given the wide applicability of NLSMs, e.g.~as the effective description of physical systems both in the continuum and on the lattice, we expect that these results will prove useful in many applications.

This letter is organised as follows. In section \ref{sec:RG_NLSM} we review NLSMs, their categorical symmetries, and their RG flow. In section \ref{sec:1-loop_general} we show that the 1-loop $\beta$-function for $G$ vanishes precisely due to the self-duality condition. The same happens at 2-loops, as we verify in section \ref{sec:2_loop}. In section \ref{sec:WZW} we argue that the non-renormalisation holds also at the non-perturbative level, based on the examples of Wess-Zumino-Witten (WZW) models. In section \ref{sec:2-loop_macnilfold} we discuss an explicit example where $M$ is a squashed 3-sphere. We finish with a discussion of our results and outlook in section \ref{sec:outlook}.

\section{Non-Linear Sigma Models and their renormalisation}\label{sec:RG_NLSM}

We begin by reviewing the necessary material regarding NLSMs, following \cite{Arias-Tamargo:2025fhv,Arias-Tamargo:2025xdd,Hull2006}. We consider the situation where the target space has a U(1) isometry without fixed points. In this case, $M$ is a fibre bundle over a base $B$,
\begin{align}\label{eq:fibration}
    S^1\hookrightarrow M \twoheadrightarrow B\,.
\end{align}

The topology of the NLSM is specified by two curvature 2-forms which we denote by $F$ and $\widetilde{F}$. The former is associated with the metric $g$ and can be written explicitly as $F=\dd\xi$ where $\xi$ is the 1-form dual to the Killing vector $k$, with components $\xi_i=G^{-1}g_{ij}k^j$. This $F$ is the curvature of the bundle \eqref{eq:fibration}. The other 2-form $\widetilde{F}$ is associated with the WZ term and is given by the contraction of $H$ with the Killing vector, $\widetilde{F}=(2\pi)\iota_k H$.

Both $F$ and $\widetilde{F}$ have the important property that they are \emph{basic}, i.e.~they do not depend on the fibre coordinate ($\mathcal{L}_kF=\mathcal{L}_k\widetilde{F}=0$) and they have no component along the fibre ($\iota_kF=\iota_k\widetilde{F}=0$). This means they can be understood as closed 2-forms on the base $B$. The normalisation is chosen such that
\begin{align}\label{eq:F_quant}
    \frac{1}{2\pi}\int_\Sigma F \in \mathbb{Z}\,,\quad \frac{1}{2\pi}\int_\Sigma \widetilde{F} \in \mathbb{Z}\,,
\end{align}
for $\Sigma$ any 2-cycle on the base $B$. The cohomology classes $[F/2\pi],\, [\widetilde{F}/2\pi]\in H^2(B,\mathbb{Z)}$ are called the \emph{Chern class} and \emph{H-class} respectively, and they determine the symmetries of the NLSM \cite{Arias-Tamargo:2025fhv}.

The geometry of the NLSM is further specified by $g$ and $H$. It will be useful to decompose them into components on the base and the fibre of \eqref{eq:fibration} as \cite{Hull2006}
\begin{align}\label{eq:base_fibre_decomp}
    g=\bar{g}+G\, \xi\otimes \xi\,,\quad H=\bar{H}+\frac{1}{2\pi}\widetilde{F}\wedge \xi\,,
\end{align}
where $\bar{g}$ and $\bar{H}$ are a metric and a 3-form on $B$.

When the topology of the NLSM is non-trivial, the condition \eqref{eq:self_duality_radius} for there to be a categorical symmetry has to be supplemented with a condition on $F$ and $\widetilde{F}$ \cite{Arias-Tamargo:2025xdd,Arias-Tamargo:2025fhv},
\begin{align}\label{eq:self_duality_full}
    G=\frac{p}{2\pi q}\,,\quad p F=q\widetilde{F}\,.
\end{align}
These are the full conditions for the NLSM to host a self-duality defect. The choice of coprime integers $p,q$ specifies the particular defect under consideration \footnote{More precisely, $p$ and $q$ specify a $\Z_p\times\Z_q$ subgroup of the global symmetry of the NLSM to be gauged in half-space. These symmetries and their 't Hooft anomalies are determined by the topology of the NLSM. The condition $\gcd(p,q)=1$ is the requirement that these discrete subgroups do not have a mixed 't Hooft anomaly. There are cases where the two individual factors can have pure 't Hooft anomalies, leading to further restrictions on $p$ and $q$ \cite{Arias-Tamargo:2025fhv}.}.

We are interested in computing the $\beta$-function for $G$. This can be extracted from the $\beta$-function of the metric $g$ which, in the 1-loop approximation, can be written \cite{Friedan:1980jf,Curtright:1984dz}
\begin{align}\label{eq:beta_function_metric}
    \beta_{ij}^{g} = \alpha' \left( R_{ij} - \frac{1}{4} H_{ikl} H\indices{_j^{kl}} \right) +O(\alpha'^2)\,,
\end{align}
where $R_{ij}$ is the target space Ricci tensor. We will denote the $n$-loop contribution to $\beta^g$ by $\beta^{g,\,(n)}$.

Isometries are preserved by the RG flow \cite{Friedan:1980jm}. This is a consequence of \eqref{eq:Lie_derivatives}, which implies that $\mathcal{L}_k\, \beta_{ij}^{g,\,(1)}=0$.  Therefore, at any scale $\Lambda$, the renormalised metric $g(\Lambda)$ will satisfy $\mathcal{L}_k g(\Lambda)=0$.  This is in fact true to all orders since $\beta^{g,\,(n)}_{ij}$ is always a tensor on $M$ constructed from $g$ and $H$. Since the isometry is preserved, the target $M$ can be written as in \eqref{eq:fibration} at all scales and the decompositions \eqref{eq:base_fibre_decomp} also hold throughout the flow, even if all quantities involved (in particular the geometry of the base $B$) are renormalised. Naïvely, one expects the radius of the $S^1$ fibre, $G^{1/2}$, to be renormalised as well. The claim of this work is that if \eqref{eq:self_duality_full} is satisfied then $G$ is \emph{not} renormalised as it is protected by the additional categorical symmetry.

We will adopt coordinates $X^i=\{Y^\mu,X\}$ on $M$ where $Y^\mu$ are coordinates on $B$ ($\mu=1,\dots,N-1$) and $X$ is a coordinate along the fibre such that $k=\partial_X$. With this choice, \eqref{eq:base_fibre_decomp} implies that $G=g_{NN}$, so its $\beta$-function is $\beta_G = \beta^g_{NN}$. We will show below that this component of \eqref{eq:beta_function_metric} vanishes when \eqref{eq:self_duality_full} is satisfied \footnote{It is not strictly necessary for the $\beta$ function to vanish, as the target space geometry would be invariant under RG if $\beta^g_{ij} = 2\nabla_{(i} V_{j)}$ for some vector $V$ \cite{Friedan:1980jf,Hull:1985rc}. In the present case, we will see that $\beta^g_{NN}$ vanishes and this technicality is not required.}.

\section{1-loop}\label{sec:1-loop_general}

Our goal now is to evaluate $\beta_G$. Our strategy will be to introduce a non-coordinate basis and exploit the Cartan structure equations together with the decompositions \eqref{eq:base_fibre_decomp} to relate the relevant components of $H$ and the Riemann tensor to $G$, $F$, and $\widetilde{F}$.

We denote the non-coordinate basis of the cotangent space of $M$ by $\{e^{\hat{a}}\}$ and the vielbein on $M$ by $e^{\hat{a}}_i$, such that $g_{ij}=e^{\hat{a}}_i e^{\hat{b}}_j \delta_{\hat{a}\hat{b}}$. Components of tensors in this basis have hatted indices $\hat{a}=(\hat{\alpha},\hat{N})$ with $\hat{\alpha}=1,\dots,N-1$ and $\hat{N}=N$, which are raised and lowered with the flat metric $\delta_{\hat{a}\hat{b}}$. Let $\bar{e}^{\hat{\alpha}}_\mu$ be a vielbein for the base $B$ such that $\bar{g}_{\mu\nu} = \bar{e}^{\hat{\alpha}}_\mu \bar{e}^{\hat{\beta}}_\nu \delta_{\hat{\alpha}\hat{\beta}}$. Equation \eqref{eq:base_fibre_decomp} implies
\begin{align}\label{eq:choice_vielbeins}
    e^{\hat{\alpha}}=\bar{e}^{\hat{\alpha}}\,,\quad e^{\hat{N}}=G^{1/2}\xi\,.
\end{align}

The $\beta$-function for $G$ can be written directly in the non-coordinate basis. We have
\begin{align}
    \beta_G = \beta^g_{NN} = G \beta^g_{\hat{N}\hat{N}}\,,
\end{align}
so in particular at 1-loop,
\begin{align}\label{eq:1-loop_G}
    \beta^{(1)}_G = \alpha' G\left(R_{\hat{N}\hat{N}}-\frac{1}{4}H_{\hat{N}\hat{a}\hat{b}}H\indices{_{\hat{N}}^{\hat{a}\hat{b}}}\right)\,.
\end{align}

We begin with the $H^2$ term. From \eqref{eq:base_fibre_decomp}, the non-zero components of $H$ in the non-coordinate basis are
\begin{equation}\label{eq:H_components}
\begin{cases}
    H_{\hat{\alpha}\hat{\beta}\hat{\gamma}} = \bar{H}_{\hat{\alpha}\hat{\beta}\hat{\gamma}}\,,\\ H_{\hat{N}\hat{\alpha}\hat{\beta}} = \frac{1}{2\pi G^{1/2}} \widetilde{F}_{\hat{\alpha}\hat{\beta}} \,,
    \end{cases}
\end{equation}
and therefore
\begin{align}\label{eq:H2}
H_{\hat{N}\hat{a}\hat{b}}H\indices{_{\hat{N}}^{\hat{a}\hat{b}}} = \frac{1}{(2\pi)^2 G}\widetilde{F}_{\hat{\alpha}\hat{\beta}}\widetilde{F}^{\hat{\alpha}\hat{\beta}}\,.
\end{align}

Evaluating the Ricci tensor requires more work. Denoting a torsion-free spin connection 1-form by $\omega\indices{^{\hat{a}}_{\hat{b}}}$, the Cartan structure equations are
\begin{gather}
    \dd e^{\hat{a}}+\omega\indices{^{\hat{a}}_{\hat{b}}}\wedge e^{\hat{b}}=0\,,\label{eq:first_cartan_eq}\\
    \mathcal{R}\indices{^{\hat{a}}_{\hat{b}}}=\dd \omega\indices{^{\hat{a}}_{\hat{b}}}+\omega\indices{^{\hat{a}}_{\hat{c}}}\wedge \omega\indices{^{\hat{c}}_{\hat{b}}}\label{eq:second_cartan_eq}\,.
\end{gather}
Here $\mathcal{R}\indices{^{\hat{a}}_{\hat{b}}}$ is the curvature 2-form (not the Ricci tensor) which is related to the Riemann tensor by
\begin{align}\label{eq:R_versus_R}
    \mathcal{R}\indices{^{\hat{a}}_{\hat{b}}}=\frac{1}{2}R\indices{^{\hat{a}}_{\hat{b}\hat{c}\hat{d}}}\,e^{\hat{c}} \wedge e^{\hat{d}} \,.
\end{align}
Inserting \eqref{eq:choice_vielbeins} into \eqref{eq:first_cartan_eq} gives the spin connection 
\begin{align}\label{eq:omega}
\begin{cases}
   \, \omega_{\hat{\alpha}\hat{\beta}} = -\omega_{\hat{\beta}\hat{\alpha}} = \bar{ \omega}_{\hat{\alpha}\hat{\beta}}-\frac{1}{2}G F_{\hat{\alpha}\hat{\beta}}\xi\,, \\
   \, \omega_{\hat{N}\hat{\beta}} = -\omega_{\hat{\beta}\hat{N}} = \frac{1}{2}G^{1/2} F\indices{_{\hat{\beta}\hat{\alpha}}}\bar{e}^{\hat{\alpha}}\,, \\
   \, \omega_{\hat{N}\hat{N}}=0\,,
\end{cases}
\end{align}
where $\bar{\omega}_{\hat{\alpha}\hat{\beta}}$ is a spin connection on the base, satisfying $\dd \bar{e}^{\hat{\alpha}}+\bar{ \omega}\indices{^{\hat{\alpha}}_{\hat{\beta}}}\wedge \bar{e}^{\hat{\beta}}=0$.

The Riemann tensor can then be found from \eqref{eq:second_cartan_eq} and \eqref{eq:R_versus_R}. In fact, since we only require the $R_{\hat{N}\hat{N}}$ component of the Ricci tensor, it suffices to compute $\mathcal{R}\indices{^{\hat{\alpha}}_{\hat{N}}}$, which is
\begin{align}
    \mathcal{R}\indices{^{\hat{\alpha}}_{\hat{N}}} &= -\frac{1}{4} G^{-1/2} F\indices{^{\hat{\alpha}}_{\hat{\beta}}} \dd{G} \wedge \bar{e}^{\hat{\beta}} - \frac{1}{2} G^{1/2} \bar{\nabla}_{\hat{\beta}} F\indices{^{\hat{\alpha}}_{\hat{\gamma}}} \bar{e}^{\hat{\beta}} \wedge \bar{e}^{\hat{\gamma}} \nonumber\\
    &- \frac{1}{4} G F\indices{^{\hat{\alpha}}_{\hat{\gamma}}}F\indices{^{\hat{\gamma}}_{\hat{\beta}}} \bar{e}^{\hat{\beta}} \wedge e^{\hat{N}} \,,\label{eq:curv2form_working2}
\end{align}
where $\bar{\nabla}$ is the Levi-Civita connection associated with the base metric $\bar{g}$. From \eqref{eq:R_versus_R} we extract the components of the Riemann tensor
\begin{align}\label{eq:RaNbc}
\begin{cases}
    \, R\indices{^{\hat{\alpha}}_{\hat{N}\hat{\beta}\hat{\gamma}}} = -\frac{1}{2} G^{1/2} \bar{\nabla}^{\hat{\alpha}} F_{\hat{\beta}\hat{\gamma}} + \frac{1}{2} G^{-1/2} (\dd{G})_{[\hat{\beta}} F\indices{_{\hat{\gamma}]}^{\hat{\alpha}}} \,, \\
    \, R\indices{^{\hat{\alpha}}_{\hat{N}\hat{\beta}\hat{N}}} = - \frac{1}{4} G F\indices{^{\hat{\alpha}}_{\hat{\gamma}}} F\indices{^{\hat{\gamma}}_{\hat{\beta}}} \,,
\end{cases}
\end{align}
which yield
\begin{align}\label{eq:Ricci}
    R_{\hat{N}\hat{N}}=\frac{G}{4} F_{\hat{\alpha}\hat{\beta}}F^{\hat{\alpha}\hat{\beta}} \,.
\end{align}
Finally, inserting \eqref{eq:H2} and \eqref{eq:Ricci} into \eqref{eq:1-loop_G},
\begin{align}\label{eq:1-loop_G_result}
    \beta^{(1)}_G = \frac{\alpha'}{4}\left(G^2 F_{\hat{\alpha}\hat{\beta}}F^{\hat{\alpha}\hat{\beta}}-\frac{1}{(2\pi)^2}\widetilde{F}_{\hat{\alpha}\hat{\beta}}\widetilde{F}^{\hat{\alpha}\hat{\beta}}\right)\,.
\end{align}
Comparing with the self-duality conditions \eqref{eq:self_duality_full}, it follows that 
\begin{align}
  \beta_G^{(1)}=0\,
\end{align}
if they are satisfied, i.e.~if the couplings are such that the NLSM admits a self-duality defect. 

Remarkably, the $\beta$-function vanishes if and only if there is a self-duality symmetry. This follows since requiring $\beta_G^{(1)}=0$ imposes $\widetilde{F} = \pm 2\pi G F$, but $F/2\pi$ and $\widetilde{F}/2\pi$ must both have integer periods so $2\pi G \in \mathbb{Q}$.

The fact that $G$ is not renormalised implies that the self-duality conditions will hold throughout the RG flow, so the categorical symmetry is present at all scales. Reversing the logic is interesting: since the theory in the UV has the categorical symmetry and this is preserved along the RG flow, the coupling $G$ is protected from quantum corrections. The cancellation of its $\beta$-function precisely expresses this fact.

\section{2-loop}\label{sec:2_loop}

We now demonstrate that the $\beta$-function for the coupling $G$ vanishes to 2-loop order when the self-duality conditions \eqref{eq:self_duality_full} are satisfied. The 2-loop $\beta$-function of the metric of NLSMs with WZ term can be written \cite{Metsaev:1987bc,Hull:1987pc,Zanon:1987pp}\footnote{The $\beta$-function is scheme dependent and here we use eq.~(32) of \cite{Metsaev:1987bc} in the $f_1=-1$ scheme.}
\begin{align}
   \beta_{ij}^{g,\,(2)}&=\frac{\alpha'^2}{2}\left(R_{iklm}R\indices{_j^{klm}}-\frac{1}{2}R\indices{_{(i}^{klm}}H_{j)kn}H\indices{_{lm}^n}\right.\nonumber\\
 & \left.-\frac{1}{2}R_{iklj}(H^2)^{kl}+\frac{1}{8}(H^4)_{ij}+\frac{1}{8}H_{ikl}H\indices{_{jm}^l}(H^2)^{km}\right.\nonumber\\  & \left.+\frac{1}{12}\nabla_i H_{klm}\nabla_j H^{klm}-\frac{1}{4}\nabla_k H_{ilm}\nabla^k H\indices{_j^{lm}}\right)\,, \label{eq:2-loop}
\end{align}
where $\nabla$ is the Levi-Civita connection for $g$ and
\begin{align}
    (H^2)_{ij}=H_{ikl}H\indices{_j^{kl}}\,,\quad (H^4)_{ij}=H\indices{^k_{l(i}}H\indices{_{j)mn}}H^{npl}H\indices{_{pk}^m}\,.
\end{align}

It is again simpler to use the non-coordinate basis, in which we have $\beta_G^{(2)}  = G \beta^{g,\,(2)}_{\hat{N}\hat{N}}$. 
The $\beta$-function can be more succinctly written in terms of the combination \footnote{This is the curvature tensor of the connection with torsion $\hat{\Gamma}\indices{_{ij}^k} = \Gamma\indices{_{ij}^k} - \frac{1}{2} H\indices{_{ij}^k}$, where $\Gamma\indices{_{ij}^k}$ are the Christoffel connection coefficients associated with $g$.}
\begin{equation}
\begin{split}\label{eq:Rhat}
    \hat{R}\indices{^{\hat{a}}_{\hat{b}\hat{c}\hat{d}}}&= R\indices{^{\hat{a}}_{\hat{b}\hat{c}\hat{d}}} - \frac{1}{2} H\indices{_{\hat{e}[\hat{c}}^{\hat{a}}} H\indices{_{\hat{d}]\hat{b}}^{\hat{e}}} - \nabla_{[\hat{c}} H\indices{_{\hat{d}]\hat{b}}^{\hat{a}}} \,,
\end{split}
\end{equation}
in terms of which the 2-loop $\beta$-function for $G$ becomes
\begin{align}
    \beta^{(2)}_{G} &= \frac{\alpha'^2 G}{2} \left( \hat{R}\indices{^{\hat{a}\hat{b}\hat{c}}_{\hat{N}}} \hat{R}_{\hat{N}\hat{a}\hat{b}\hat{c}} - \frac{1}{2} \hat{R}\indices{^{\hat{b}\hat{c}\hat{a}}_{\hat{N}}} \hat{R}_{\hat{N}\hat{a}\hat{b}\hat{c}} \right.\nonumber\\
    &\left. + \frac{1}{2} \hat{R}_{\hat{a}\hat{N}\hat{N}\hat{b}} (H^2)^{\hat{a}\hat{b}} \right) \,. \label{eq:beta2_Rhat}
\end{align}

Computing the relevant components of $\hat{R}$ in \eqref{eq:beta2_Rhat} requires the components  \eqref{eq:RaNbc}  of the Riemann tensor
as well as
\begin{equation}\label{eq:nabla_H}
\begin{cases}
    \nabla_{\hat{\gamma}}H_{\hat{\alpha}\hat{\beta}\hat{N}} = G^{1/2} \left( \frac{1}{2\pi G}\bar{\nabla}_{\hat{\gamma}} \widetilde{F}_{\hat{\alpha}\hat{\beta}} + \frac{1}{2} F\indices{^{\hat{\delta}}_{\hat{\gamma}}} \bar{H}_{\hat{\alpha}\hat{\beta}\hat{\delta}} \right) \,, \\
    \nabla_{\hat{N}} H_{\hat{\alpha}\hat{\beta}\hat{\gamma}} = \frac{3}{2} G^{1/2} F\indices{^{\hat{\delta}}_{[\hat{\alpha}}} \bar{H}_{\hat{\beta}\hat{\gamma}]\hat{\delta}} \,.
\end{cases}
\end{equation}
These can be derived using the components of the spin connection \eqref{eq:omega} and the fact that $\mathcal{L}_k \bar{H}=0$. We have set $\dd G=0$ in \eqref{eq:nabla_H} to simplify their form since we will ultimately impose the conditions \eqref{eq:self_duality_full} which fix $G$ to a particular constant.

Now we can insert \eqref{eq:H_components}, \eqref{eq:RaNbc} and \eqref{eq:nabla_H} into \eqref{eq:Rhat}. We find,
\begin{align}\label{eq:Rhat_aNbN}
\begin{cases}
    \hat{R}\indices{_{\hat{\alpha}\hat{N}\hat{\beta}\hat{N}}} = -\frac{1}{4} G \left( F\indices{_{\hat{\alpha}\hat{\gamma}}} F\indices{^{\hat{\gamma}}_{\hat{\beta}}} - \frac{1}{(2\pi G)^2} \widetilde{F}\indices{_{\hat{\alpha}\hat{\gamma}}} \widetilde{F}\indices{^{\hat{\gamma}}_{\hat{\beta}}} \right) \,, \\
    \hat{R}_{\hat{\alpha}\hat{\beta}\hat{\gamma}\hat{N}} = -\frac{1}{2} G^{1/2} \left( \bar{\nabla}_{\hat{\gamma}} \mathcal{F}_{\hat{\alpha}\hat{\beta}} + \mathcal{F}\indices{^{\hat{\delta}}_{[\hat{\alpha}}} \bar{H}_{\hat{\beta}]\hat{\gamma}\hat{\delta}} \right) \,,
\end{cases}
\end{align}
where
\begin{align}
 \mathcal{F}_{\hat{\alpha}\hat{\beta}} = F_{\hat{\alpha}\hat{\beta}} - \frac{1}{2\pi G}\widetilde{F}_{\hat{\alpha}\hat{\beta}}\,.   
\end{align}

When the conditions \eqref{eq:self_duality_full} are imposed $\mathcal{F}=0$, and both components in \eqref{eq:Rhat_aNbN} vanish; then \eqref{eq:beta2_Rhat} gives
\begin{align}
    \beta_G^{(2)}=0\,.
\end{align}
That is, $G$ is not renormalised in the 2-loop approximation when the NLSM admits a self-duality defect. Again, the inverse implication is also true: the vanishing of $\beta_G^{(1)}$ implies the existence of the categorical symmetry, which in turn implies the vanishing of $\beta_G^{(2)}$.
We note that the $\hat{R}_{\hat{\alpha}\hat{\beta} \hat{\gamma} \hat{\delta}}$ components are generically non-zero even when \eqref{eq:self_duality_full} are satisfied, and include contributions from the curvature tensor of the base metric $\bar{g}$, and also from $\bar{H}$. They determine the RG flow of the other components of the metric so, while $G=g_{NN}$ is protected by the categorical symmetry, the NLSM is not generically conformal.

\section{Non-perturbative -- WZW Models}\label{sec:WZW}

Next, consider a NLSM on a target space which is a compact group manifold, $\mathcal{G}$. The action can be written
\begin{align}\label{eq:WZW_action}
    S=-\frac{1}{4\pi\alpha'\lambda^2}\int_W \Tr\left(g^{-1} d g\right)^2+\frac{\kappa}{24\pi^2\alpha'}\int_V \Tr\left(g^{-1}d g\right)^3\,,
\end{align}
where $g: W\to \mathcal{G}$ parametrises the dynamical fields. Since $\kappa\in\Z$ is quantised and does not get renormalised, this model has only one independent running coupling $\lambda$. Moreover, this remains true at all scales due to the $(\mathcal{G}\times \mathcal{G})/Z(\mathcal{G})$ isometry. In \cite{Arias-Tamargo:2025xdd} the cases $\mathcal{G}=\text{SU}(n), \,\text{Spin}(n)$ were studied and it was found that the self-duality conditions \eqref{eq:self_duality_full} become
\begin{align} \label{eq:duality_WZW}
    \lambda^2 = \frac{4\pi}{\kappa}\,.
\end{align}
The same logic of the previous sections then implies that \eqref{eq:duality_WZW} should be preserved under RG flow. Since $\lambda$ is the only independent running coupling, it follows that the full theory would be scale invariant if the $\beta$-function for $\lambda$ vanishes. Indeed, at the value \eqref{eq:duality_WZW} of the coupling, the action \eqref{eq:WZW_action} describes the WZW model at level $\kappa$, which is exactly conformal \cite{Witten:1983ar}. That is,
\begin{align}
    \beta_{\lambda^2}=0\,,
\end{align}
at the self-dual value \eqref{eq:duality_WZW}. From the point of view of the present work, we interpret this as a consequence of the fact that the categorical symmetry protects the coupling $G=2/\lambda^2$ from all quantum corrections, both perturbative and non-perturbative. As $\lambda$ is the only coupling constant, this then  implies conformal symmetry.

\section{Example -- Squashed 3-sphere}\label{sec:2-loop_macnilfold}

We now study a concrete example in which the target space is topologically an $S^3$, but with an extra squashing parameter $\ell$ that modifies the radius of the fibre. We also include a WZ term, so that the total action is
\begin{align}\label{eq:squash_squash}
    S=\frac{1}{2\pi \alpha'}&\left(\frac{r_0^2}{8}\int_W \left(d\theta^2+\sin^2\theta \,d\psi^2+4\ell^2(A+d\phi)^2\right)\right.\nonumber\\
    &+\frac{\kappa}{4\pi}\int_V \sin\theta\,d\theta\wedge d\psi\wedge d\phi\bigg)\,.
\end{align}
The $S^2$ base is parametrised by the angles $\theta\in[0,\pi]$ and $\psi\in[0,2\pi]$, the Hopf fibre by $\phi\in[0,2\pi]$, and $\kappa \in\mathbb{Z}$. Denoting an open cover of the $S^2$ base by $\{U_+,U_-\}$ where $U_+$ excludes $\theta=\pi$ and $U_-$ excludes $\theta=0$, the connection can be written in each patch as
\begin{align}
    A_\pm =\frac{1}{2}\left( \pm1 - \cos\theta\right)\,\dd\psi\,.
\end{align}
This space has an isometry generated by the Killing vector $k=\partial_\phi$ whose dual 1-form is $\xi=A+\dd\phi$. The Chern and H-classes are
\begin{align}
    \frac{1}{2\pi}\int_{S^2}F=1\,,\quad\frac{1}{2\pi}\int_{S^2}\widetilde{F}=\kappa\,.
\end{align}
The norm squared of the Killing vector is $G=r_0^2\ell^2$, and the self-duality conditions \eqref{eq:self_duality_full} become
\begin{align}\label{eq:self_duality_squashed}
    r_0^2\ell^2=\frac{\kappa}{2\pi}\,.
\end{align}
From section \ref{sec:RG_NLSM}, the $\beta$-function for $r_0^2\ell^2$ corresponds to the $g_{\phi\phi}$ component of the metric. We can compute it by directly evaluating \eqref{eq:beta_function_metric} and \eqref{eq:2-loop} for the model \eqref{eq:squash_squash},
\begin{align}           
\beta_{r_0^2\ell^2}^{(1)}&=2\alpha'\left(\ell^4-\frac{\kappa^2}{4\pi^2 r_0^4}\right)\,,\\
    \beta_{r_0^2\ell^2}^{(2)} &= \frac{\alpha'^2}{2}\left(\frac{2\ell^6}{r_0^2}-\frac{3\kappa^2\ell^2}{\pi^2 r_0^6}+\frac{5\kappa^4}{8\pi^4 r_0^{10}}\right) \,.
\end{align}
For the particular values \eqref{eq:self_duality_squashed}, indeed we verify that
\begin{align}
   \beta_{r_0^2\ell^2}^{(1)}=0\,,\quad \beta_{r_0^2\ell^2}^{(2)}=0\,.
\end{align}
As a result, the coupling which gives the radius of the $S^1$ fibre is not renormalised in the 2-loop approximation when the categorical symmetry is present. The fact that the coupling \eqref{eq:self_duality_squashed} is preserved to 2-loop order in this model has been noted in \cite{Schubring:2020uzq}, where it was conjectured that this should hold at all loop orders (see also \cite{Levine:2021fof}). As discussed in section~\ref{sec:WZW}, there is compelling evidence that the global categorical symmetry should protect the coupling \eqref{eq:self_duality_squashed} against any renormalisation, giving support to their conjecture. We remark that the other components of the metric are renormalised and the theory is not conformal, even when \eqref{eq:self_duality_squashed} is satisfied. In particular, at the self-dual values \eqref{eq:self_duality_squashed} we find
\begin{align}\label{eq:other_betafnct_sphere}
    \beta_{\theta\theta}^{g,\,(1)}&= \frac{1}{8} \beta_{r_0^2}^{(1)} = \alpha'(1-\ell^2) = \alpha' \left( 1 - \frac{\kappa}{2\pi r_0^2} \right) \,,
\end{align}
so both $r_0$ and $\ell$ individually run, with $r_0 \ell$ fixed. At low energies, the flow drives $\ell\to1$ and the theory approaches the SU(2)$_\kappa$ WZW model. The self-duality symmetry is present at all scales.

There are many other examples of NLSMs that host a self-duality defect, some of which were discussed in \cite{Arias-Tamargo:2025xdd}. Non-conformal ones include particular deformations of Lens spaces, certain nilmanifolds (non-trivial torus bundles over tori), and $T^{p,q}$ manifolds \cite{Romans:1984an}\footnote{$T^{p,q}$ manifolds are $S^1$ bundles over $S^2\times S^2$, with $p,\,q$ specifying the Chern class. There is a particular limit of the NLSM on $T^{1,q}$ with WZ term where the target space factorises into a squashed $S^3$ and a round $S^2$ \cite{Levine:2021fof}. The former is the example studied in detail above.}. In all these cases, the radius of the $S^1$ fibre is protected from quantum corrections by the categorical symmetry. 

\section{Conclusion and outlook}\label{sec:outlook}

In this work we have argued that in two-dimensional NLSMs with a categorical self-duality symmetry there is a coupling that is protected from all quantum corrections. Our evidence for this claim is that 1) the 1- and 2-loop $\beta$-function for this coupling vanishes in these models if and only if they admit the self-duality defect, and 2) when the target space is a group manifold as in section \ref{sec:WZW}, the fact that WZW models are conformal can be interpreted as a particular instance of this non-renormalisation result which is valid non-perturbatively. In cases where the NLSM under study is not conformal, such an explicit check of the non-renormalisation at the non-perturbative level is not available at present. Nevertheless, we believe that the fact that there is a symmetry argument supporting it suggests the result can be applied in a much broader context. In particular, we believe it can be applied if the NLSM is strongly coupled in the IR, so that the coupling $G$ will not be renormalised.

One of the main directions for future research is, of course, to gather further evidence for the validity of this argument in non-conformal theories. In this context, it would prove useful to study theories in which the RG flow is more constrained, e.g.~supersymmetric NLSMs or integrable deformations which preserve the symmetry. 

Another intriguing direction is the apparent relation between the categorical symmetry and integrability. There are cases where the presence of the self-duality defect coincides with the condition for the NLSM to be classically integrable. This occurs, for instance, for the $T^{1,q}$ spaces studied in \cite{Levine:2021fof}. One consequence is that the condition for classical integrability is protected under RG flow by the categorical symmetry in these examples. It would be interesting to better understand the relation between the presence of these extra symmetries and the integrability of these models.

In this work, we have focused on NLSMs whose target space has a single U(1) isometry. The construction of the non-invertible defect can be extended to examples where the relevant isometry is instead U(1)$^d$, and in these cases the self-duality conditions involve a $d\times d$ matrix of couplings \cite{Arias-Tamargo:2025xdd}. An analogous argument to the one presented here then suggests that there may be many relations between couplings which will be protected from quantum corrections in theories where such categorical symmetries are present, putting multiple constraints on the RG flow. It would also be interesting to verify this explicitly. 

There are several interesting consequences of the fact that these self-duality symmetries are present throughout the RG flow of non-conformal NLSMs. First, at low energies the effective field theory describing the IR dynamics of the NLSM should either preserve or spontaneously break the symmetry. In the context of the Landau paradigm, this would allow for an understanding of the phases of the theory which can be distinguished by the realisation of the categorical symmetry \cite{Bhardwaj:2023fca}. Second, scattering amplitudes of the QFT will be subject to Ward identities derived from the categorical symmetry \cite{Copetti:2024rqj,Copetti:2024dcz} (see also \cite{Frolov:2025ozz,Shimamori:2025ntq}). We hope that these insights, together with the non-renormalisation of the coupling, can be combined to substantially advance our understanding of physical systems described by NLSMs.

Self-duality symmetries are not unique to two dimensions \cite{Kaidi:2021xfk,ChoiCordovaHsinLam2021}. In fact, in four dimensions they are quite common even in interacting theories, e.g.~in the context of the class $\mathcal{S}$ construction \cite{Bashmakov:2022uek,Antinucci:2022cdi,Carta:2023bqn}. It is then a natural question to ask if similar results to the ones we present here hold also in higher dimensions.\\

\noindent\textbf{Acknowledgements.} It is a pleasure to thank Giovanni Galati, Shota Komatsu, Arkady Tseytlin, and Anders Wallberg for interesting discussions. GAT and CMH are supported by the STFC Consolidated Grant ST/X000575/1. MVCH is supported by a President's Scholarship from Imperial College London.

\bibliography{refs}

\end{document}